\newif\ifsubmode
\submodefalse
\submodetrue

\documentclass{emulateapj}

\ifsubmode
\else
\slugcomment{\today}
\fi

\def\rb{{\bf r}}                   

\def\kms{\hbox{$~$km$~$s$^{-1}$}}

\def\snid{\ifmmode{\rm \tt SNID}\else{\tt SNID}\fi}
\def\dm15{\ifmmode{\Delta m_{15}}\else{$\Delta m_{15}$}\fi}
\def\magarcsec2{\ \rm{mag
\ arcsec}^{-2}}

\shorttitle{Spectral Identification of SNe with Light Echoes}
\shortauthors{Rest et al.}

\begin{document}

\title{Spectral Identification of an Ancient Supernova using Light Echoes in the LMC}

\author{A. Rest\altaffilmark{1,2,3}, T. Matheson\altaffilmark{4}, 
S. Blondin\altaffilmark{5}, M. Bergmann\altaffilmark{6},
D. L. Welch\altaffilmark{7}, N. B. Suntzeff\altaffilmark{8},
R. C. Smith\altaffilmark{1}, K. Olsen\altaffilmark{1},
J. L. Prieto\altaffilmark{9}, A. Garg\altaffilmark{2},
P. Challis\altaffilmark{5}, C. Stubbs\altaffilmark{2,5},
M. Hicken\altaffilmark{5}, M. Modjaz\altaffilmark{5},
W. M. Wood-Vasey\altaffilmark{5}, A. Zenteno\altaffilmark{1},
G. Damke\altaffilmark{1}, A. Newman\altaffilmark{10},
M. Huber\altaffilmark{11}, K. H. Cook\altaffilmark{11},
S. Nikolaev\altaffilmark{11}, A. C. Becker\altaffilmark{12},
A. Miceli\altaffilmark{12}, R. Covarrubias\altaffilmark{12,13},
L. Morelli\altaffilmark{14}, G. Pignata\altaffilmark{14,15},
A. Clocchiatti\altaffilmark{14}, D. Minniti\altaffilmark{14}, and 
R. J. Foley\altaffilmark{16}}

\altaffiltext{1}{Cerro Tololo Inter-American Observatory (CTIO), Colina el Pino S/N, La Serena, Chile}

\altaffiltext{2}{Physics Department, Harvard University, 17 Oxford Street, Cambridge, MA 02138}

\altaffiltext{3}{Goldberg Fellow}

\altaffiltext{4}{National Optical Astronomy Observatory, 950 N. Cherry Ave., Tucson, AZ 85719-4933}

\altaffiltext{5}{Harvard-Smithsonian Center for Astrophysics, 60 Garden St., 
Cambridge, MA 02138.} 

\altaffiltext{6}{Gemini Observatory, Casilla 603, La Serena, Chile}

\altaffiltext{7}{Dept. of Physics and Astronomy, McMaster University,
Hamilton, Ontario, L8S 4M1, Canada}

\altaffiltext{8}{Dept. of Physics, Texas A\&M University, College Station, TX 77843-4242}

\altaffiltext{9}{Dept. of Astronomy,
Ohio State University, 140 West 18th Ave., Columbus, OH 43210-1173}

\altaffiltext{10}{Dept. of Physics, Washington University, 1 Brookings Drive, St. Louis, MO 63130}

\altaffiltext{11}{Lawrence Livermore National Laboratory, 7000 East Ave., Livermore, CA 94550}

\altaffiltext{12}{Dept. of Astronomy, University of Washington, Box 351580, Seattle, WA 98195}

\altaffiltext{13}{Las Campanas Observatory (OCIW), Colina El Pino, Casilla 601, La Serena, Chile}

\altaffiltext{14}{Dept. of Astronomy, Pontificia Universidad Cat\'olica de Chile, Casilla 306, Santiago 22, Chile}

\altaffiltext{15}{Departamento de Astronomía, Universidad de Chile, Casilla 36-D, Santiago,
Chile}

\altaffiltext{16}{Department of Astronomy, 601 Campbell Hall, University of
California, Berkeley, CA 94720-3411}

 
\begin{abstract}
We report the successful identification of the type of the supernova
responsible for the supernova remnant SNR 0509-675 in the Large
Magellanic Cloud (LMC) using Gemini spectra of surrounding light
echoes. The ability to classify outbursts associated with centuries-old
remnants provides a new window into several aspects of supernova
research and is likely to be successful in providing new constraints
on additional LMC supernovae as well as their historical counterparts
in the Milky Way Galaxy (MWG).
The combined spectrum of echo light from SNR 0509-675 shows broad
emission and absorption lines consistent with a supernova (SN)
spectrum.  We create a spectral library consisting of 26 SNe~Ia and 6
SN~Ib/c that are time-integrated, dust-scattered by LMC dust, and
reddened by the LMC and MWG.  We fit these SN templates to the
observed light echo spectrum using $\chi^2$ minimization as well as
correlation techniques, and we find that overluminous 91T-like SNe~Ia
with $\dm15<0.9$ match the observed spectrum best.

\end{abstract}




\section{Introduction}
\label{sec:intro}

Over 100 years ago, a rapidly expanding nebula was photographed by
Ritchey around Nova Persei 1901
\citep{Ritchey01b,Ritchey01a,Ritchey02}, and it was interpreted as a
light echo from the nova explosion \citep{Kapteyn02}.  Later modelling
of the physics of the scattering and the geometry that lead to
apparent superluminal expansion confirmed this interpretation
\citep{Couderc39}.  Since then, light echoes (whereby we mean a simple
scattering echo rather than fluorescence or dust re-radiation) have
been seen in the Galactic Nova Sagittarii 1936 \citep{Swope40} and the
eruptive variable V838 Monocerotis \citep{Bond03}. Echoes have also
been observed from extragalactic supernovae, with SN~1987A being the
most famous case \citep{Crotts88,Suntzeff88,Newman06}, but also including
SNe~1991T
\citep{Schmidt94,Sparks99}, 1993J \citep{Sugerman02,Liu03}, 1995E
\citep{Quinn06}, 1998bu \citep{Cappellaro01}, 2002hh \citep{Welch07}, and 2003gd
\citep{Sugerman05,VanDyk06}.

By simple scaling arguments based on the visibility of Nova Persei
\citep{Shklovskii64,vandenBergh65,vandenBergh75}, 
light echoes from supernovae as old as a few hundred to a thousand
years can be detected, especially if the illuminated dust has regions
of high density ($\geq 10^{-8} \rm{cm}^{-3}$).  More sophisticated
models of scattered light echoes have been published
\citep{Chevalier86,Sugerman03,Patat05} but the tabulations do not
predict late-time light echo surface brightness.

The few targeted surveys for light echoes from supernovae
\citep{vandenBergh66,Boffi99} and novae
\citep{vandenBergh77,Schaefer88} have been unsuccessful. However,
these surveys did not use digital image subtraction techniques to
remove the dense stellar and galactic backgrounds. Even the bright
echoes near SN~1987A \citep{Suntzeff88} at $V\approx 21.3\magarcsec2$
are hard to detect relative to the dense stellar background of the
Large Magellanic Cloud (LMC).

During the five observing seasons allocated to SuperMACHO Project
observations, the LMC was observed repeatedly using the Mosaic imager
at the Cerro Tololo Interamerican Observatory (CTIO) Blanco 4m
telescope. An automated image reduction pipeline performed
high-precision difference-imaging from September 2001 to December
2005.  We discovered light echo systems associated with three ancient
SNe in the LMC. The echo motions trace back to three of the youngest
supernova remnants (SNRs) in the LMC: SNR 0519-69.0, SNR 0509-67.5,
and SNR 0509-68.7 (N103B).  These three remnants have also been
identified as Type Ia events, based on the X-ray spectral abundances
\citep{Hughes95}. We have dated these echoes to events 400-800 years
ago using their position and apparent motion \citep{Rest05b}. Such
light echo systems provide the extraordinary opportunity to study the
spectrum of the light from SN explosions that reached Earth hundreds
of years ago, determine their spectral types, and compare them to now
well-developed remnant structures and elemental residues.

We have obtained spectra of light echoes from each of the three light
echo groups with the Gemini-South GMOS spectrograph.  While the light
echoes of SNR 0519-69.0 and 0509-68.7 are in very crowded regions of
the LMC bar, the light echo features associated with SNR 0509-67.5 
are in a much less confused area. They are also the brightest light echo 
features we have discovered to date. We have extracted the light echo 
spectrum associated with SNR 0509-67.5 applying standard reduction 
techniques. 

The stellar spectral LMC background cannot be completely removed from
the fainter light echo features of SNR 0519-69.0 and 0509-68.7. We have
obtained multi-object spectra with using GMOS on Gemini South separated 
in time by one year in order to subtract off the stationary stellar
spectral background and retain the (apparently moving) supernova echo light.
In this paper we discuss the analysis of the light echo
spectrum of SNR 0509-67.5, and we determine the SN spectral type. We
defer the analysis of the light echo spectrum of SNR 0519-69.0 and
0509-68.7 to a future paper.

\section{Observations \& Reductions}
\label{sec:observations}

\subsection{Data Reduction of Imaging Observations}

The SuperMACHO Project microlensing survey monitored
the central portion of the LMC with a cadence of every other night 
during the five fall observing seasons beginning with September 2001.
The CTIO 4m Blanco telescope with its  8Kx8K MOSAIC imager and
atmospheric dispersion corrector were used to cover a mosaic 
of 68 pointings in an approximate rectangle $3.7\deg$ by $6.6\deg$ 
aligned with the LMC bar. The images are taken
through our custom ``VR'' filter ($\lambda_c=625\ \rm{nm}$, 
$\delta \lambda=220\ \rm{nm}$, NOAO Code c6027) with exposure times of 
60 s to 200 s, depending on the stellar densities. We used an automated 
pipeline to subtract PSF-matched template images from the most recently
acquired image to search for variability \citep{Rest05a,Garg07,Miknaitis07}. 
The resulting difference images are remarkably clean of the (constant) stellar 
background and are ideal for searching for variable objects. Our pipeline 
detects and catalogs the variable sources. 

While searching for microlensing events in the LMC, we detected groups 
of light echoes pointing back to three SNRs in the LMC, SNR 0519-69.0, 
SNR 0509-67.5, and SNR 0509-68.7 (N103B). The surface brightnesses of 
the light echoes ranged from $22.5\magarcsec2$ to the detection limit 
of the survey of about $24\magarcsec2$.

\subsection{Spectroscopic Observations}

Images were obtained on UT 2005 September 7 in the $r'$ band using the
Gemini-S GMOS spectrograph covering a 5.5 $\times$ 5.5 arcmin field
centered on the brightest echoes associated with SNR 0509-675. 
These pre-images were used to design a focal plane mask which included 
slitlets on 9 echoes, 9 stars, and 28 blank sky regions. The slits were 
1.0 arcsec wide. Spectroscopy was obtained using GMOS with the R400 grating,
yielding a resolution of 0.8\AA\ and a spectral range of 4500-8500\AA.

Spectroscopic observations were obtained on UT 2005 Nov 7,  2005 Dec 6
and 2005 Dec 7.  A total of six hour-long observations were made.  The
data were taken using the nod-and-shuffle technique
\citep{Glazebrook01}, with the telescope nodded between the on source
position and a blank sky field located 4 degrees away (off the LMC)
every 120 seconds. In total, 3 hours were spent integrating on-source. 
CuAr spectral calibration images and GCAL flatfields were interspersed 
with the science observations. The nod-and-shuffle technique provides 
for the best possible sky subtraction despite strong fringing in the 
red for the CCDs.

\subsection{Data Reduction of Spectroscopic Observations}

The GMOS data were processed using IRAF\footnote{IRAF is distributed
by the National Optical Astronomy Observatory, which is operated by
the Association of Universities for Research in Astronomy, Inc., under
cooperative agreement with the National Science Foundation.} and the
Gemini IRAF package.  The GMOS data was written as multi-extension FITS
files with three data extensions, one for each of the three CCDs in
the instrument.  We performed the initial processing by extension
(that is, by CCD), waiting until after nod and shuffle subtraction to
mosaic the extensions into a single array.  First, an overscan value
was subtracted and unused portions of the array were trimmed.  As the
CCD was dithered after each nod and shuffle observation in order to
reduce the effects of charge traps, a separate flat field was obtained
for each science observation.  The flat field images were fit with a
low-order spline for normalization and then each science frame was
flattened with the appropriate normalized flat. Both ``A'' and ``B''
nods in a given frame were flattened with the same flat. The Gemini
IRAF package \texttt{gnscombine} was used to combine all the
observations of a given light echo while also performing the
subtraction of the nod and shuffle components. The individual
extensions were then mosaicked into a single array.

Since the slitlets were oriented at various position angles tangent to
the bright portions of the echo, each individual slitlet had to be
rectified using a geometric transformation derived from the CuAr
calibration lamp spectra. The slitlets with the brightest echo
features were extracted by collapsing each slitlet along the spatial
dimension. Wavelength calibration provided by the CuAr calibration
lamps were taken through the same mask. A low-order polynomial was fit
to the calibration lamp spectra and the solution then applied to each
slitlet.  The three brightest individual slitlets could then be
combined\footnote{Each slitlet was in a different physical location on
the mask, so each spectrum has a slightly different central
wavelength, necessitating wavelength calibration before combination.}.
A slitlet that had been purposely placed in an apparently blank
portion of the field was then used to create a background spectrum to
represent the diffuse spectral background of the LMC. This background
was scaled and subtracted from the one-dimensional combined spectrum
of the echo.  Despite these efforts, some residual spectral
contamination remains, as evidenced by narrow emission lines in the
final spectrum.

\begin{figure}[t]
\epsscale{0.9}
\ifsubmode
\plotone{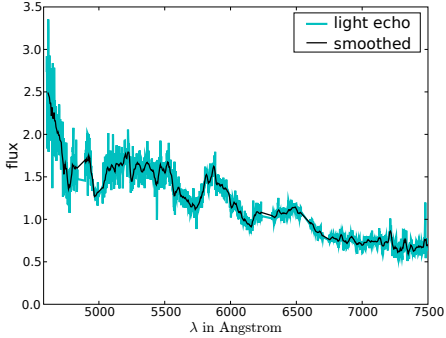}
\fi
\caption[]{
Observed spectrum of a light echo assicated with SNR 0509-675 (cyan
line). We have removed the chip gaps (4838\AA-4894\AA,
6235\AA-6315\AA), and various LMC background emission lines ([O III],
H$\alpha$, [N II], [S II]).  The black line indicates the
boxcar-smoothed spectrum. The broad absorption and emission lines are
consistent with SN spectra.
\label{fig:observedspec}}
\end{figure}

We used the spectrophotometric standard LTT~4364 \citep{Hamuy92,
Hamuy94} to flux-calibrate the individual spectra using our own
routines in IDL.  The standard star was not observed on the same
nights as the echo observations.  The relative spectrophotometry is
expected to be good at the 5\% level based on our extensive experience
with the reduction of a large sample of low-z Type Ia SNe for which we
have photometry and spectroscopy\citep{Matheson07}. We also used
LTT~4364 to remove telluric features from the spectra using techniques
described by \citet{wade88, bessell99, matheson00}.
Figure~\ref{fig:observedspec} shows the reduced spectrum of the light
echo associated with SNR 0509-675, which exhibits broad emission and
absorption features consistent with SN spectra.

\section{Analysis}
\label{sec:analysis}

Several authors have addressed the scattering of light off dust
particles and its effect on the surface brightness and spectrum of the
resulting light echoes
\citep[e.g.]{Couderc39,Chevalier86,Emmering89,Sugerman03,Patat05,Patat06}.
The light echo spectrum can be described by the time-integrated SN
spectrum attenuated by the scattering
dust \citep{Sugerman03,Patat06}. First, we describe how we create our
library of 26 time-integrated SN~Ia spectra and 6 time-integrated
SN~Ib/c spectra (see Section~\ref{sec:timeintspec}). Since our goal is
to compare SN spectra, we introduce and describe methods to correlate
and compare SN spectra in Section~\ref{sec:corrmethods}, and we test
these methods by comparing the integrated SN~Ia template spectra. We
introduce a technique that estimates the \dm15\ of a given SN~Ia based
on its correlation with other SNe~Ia. In Section~\ref{sec:scatter} we
describe how the spectrum is attenuated by the scattering dust. We
then test in Section~\ref{sec:98bu} our method to estimate the \dm15
of a SNe by correlating its light echo with dust-scattered,
time-integrated SNe~Ia template on the example of SN~1998bu.

\subsection{Time-integrated Spectra}
\label{sec:timeintspec}

The observed light echo is derived from scattering of the supernova
light off of dust sheets.  These dust sheets have light time travel
dimensions which are significant with respect to the duration of a
supernova's pre-nebular phase.  Thus, the light echo represents a
time-integration of the supernova flux attenuated by the scattering
dust \citep{Sugerman03,Patat06}.  We create 26 time-integrated SN~Ia
spectra using the lightcurve and spectral library of
\cite{Matheson07,Jha06}, and 6 time-integrated SN~Ib/c using the data
references shown in Column~(10) in Table~\ref{tab:snoverviewIbc}. We
cannot use a simple integration algorithm, like the trapezoidal rule,
since there are often significant gaps of sometimes up to 15 days in
coverage due to bright time, weather conditions, and schedule
constraints. Simple interpolation over non-homogenous coverage would
not correctly account for the non-linear shape of the SNe
lightcurves. Thus we calculate for each day a spectrum as the weighted
average of its two closest in time input spectra and scale it so that
the spectrum convolved with the $V$ filter agrees with the magnitude
from the lightcurve fit. Then we use the trapez rule to integrate.
All these steps are simple weighted summations of the input spectra,
and can be combined to a weighted sum of all the input spectrum.
Since we have normalized the input spectrum so that the spectra
convolved with the $V$ filter have the same reference magnitude
$V=15$, these weights have the added benefit that they indicate the
contribution of each input spectrum to the final integrated
spectrum. This gives us the means to test if one or two input spectra
dominate the integrated spectrum due to imperfect coverage: If the
maximum weight is large, it is an indication of such problems. We use
this maximum weight as a tool to grade the integrated spectra. In
detail, we perform the following steps on each SN:
\begin{itemize}
\item The spectral templates $\hat{S}_i$ at epochs $\hat{t}_i$ 
are normalized so that the spectra convolved with the $V$ filter have
the same reference magnitude $V=15$.
\item We fit $V$-band lightcurve templates of SN~Ia \citep{Prieto06}
and SN~Ib/c (P. Nugent online library\footnote{\texttt{http://supernova.lbl.gov/$\sim$nugent/nugent\_templates.html}}) to the observed lightcurves.
These fitted lightcurves range from -15 days to +85 days with respect
to the B-band maximum.
\item For a given time $t_k$, we find the spectra that are closest to this date 
in both time directions, $\hat{t}^-_{i1}$ and $\hat{t}^+_{i2}$. We estimate the
spectrum as the time-weighted average of these two spectra with
$c_{i1}(t_k)=(\hat{t}^-_{i2}-t_k)/(\hat{t}^+_{i2}-\hat{t}^-_{i1})$ and
$c_{i2}(t_k)=(t_k-\hat{t}^-_{i1})/(\hat{t}^+_{i2}-\hat{t}^-_{i1})$. 
For times before the
first or after the last spectrum, we just use the first and last
spectrum, respectively.  Then we normalize the spectrum by $n(t_k)$ so
that the spectrum convolved with the $V$ filter agrees with the
magnitude from the lightcurve fit.
\begin{eqnarray}
S(t_k) = n(t_k) * (c_{i1}(t_k)  \hat{S}_{i1} + c_{i2}(t_k)  \hat{S}_{i2})
\end{eqnarray}
\item We integrate the spectrum from -15 days to +85 days with respect to the 
$B$ band maximum.  Calculating the integrated spectrum is thus just a
linear combination of the input template spectra $\hat{S}_{i}$
\begin{eqnarray}
F(\lambda)  & = & \sum_k S(t_k)\\
& = & \sum_j w_j \hat{S}_{j}
\end{eqnarray}
where the $w_j$ are functions of $n(t_k)$ and $c_i(t_k)$.
\end{itemize}
As outlined before, the dominant complication creating the
time-integrated spectra is that for a given SN, the epochs for which
spectra are available are non-homogenous, and one or two spectra can
end up dominating the integrated spectrum.  If the maximum weight
$w_{max}=maximum(w_j)$ is large, it is an indication of such problems.
To reflect the relative quality of this effect, we grade our
integrated spectra by requiring that for grade $A$, $B$, and $C$ the
maximum weight fulfills $w_{max} \leq 25$\%, $w_{max} \leq 35$\%, and
$w_{max} \leq 45$\%, respectively. We find 13, 7, and 8 SNe~Ia of
grade $A$, $B$, and $C$, respectively (see
Table~\ref{tab:snoverviewIa}). For the SNe of type Ib/c, we find 3, 2,
and 1 of grade $A$, $B$, and $C$, respectively (see
Table~\ref{tab:snoverviewIbc}).

\subsection{Methods to Compare and Correlate Spectra}
\label{sec:corrmethods}

Our ultimate goal is to find the time-integrated template spectra that
match best the observed light echo spectrum in order to determine the
(sub)type of the SNe. One possibility is to fit the template spectra
to the observed light echo spectrum by a simple normalization and
calculate the $\chi^2$. The intrinsic problem with a
$\chi^2$-minimization fit of spectra is that already small but low
spatial frequency errors such as those due to errors in dereddening or
background subtraction can warp the spectrum and lead to a bad measure
of fit. An alternative to the $\chi^2$-minimization approach is the
cross-correlation technique. In this paper, we use an implementation
of the correlation techniques of \citet{Tonry79}, the SuperNova
IDentification code (\snid;
\citealt{Blondin06,Blondin07}). 
We compare and correlate the
template SN Ia spectra: Do the templates correlate better if they have
similar $\dm15$? Can we determine the $\dm15$ of a SNe template just
by correlating it to the other SNe Ia templates?  In the following we
discuss the advantages and disadvantages of these techniques.

\subsubsection{$\chi^2$-minimization}
\label{sec:chi2min}

\begin{figure}[t]
\epsscale{0.9}
\ifsubmode
\plotone{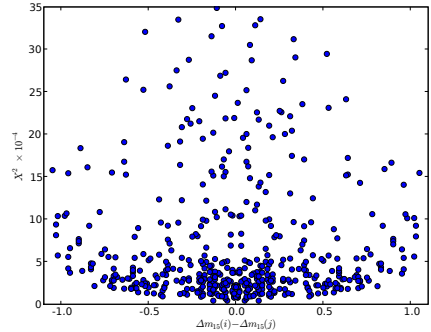}
\fi
\caption[]{This figure shows the $\chi^2$ versus their
difference in $\dm15$: $\Delta m_{15}(i) - \Delta m_{15}(j)$ for all
SN Ia in category $A$ and $B$. There is a trend that for smaller
differences in $\dm15$ the $\chi^2$ is smaller, indicating that indeed
SN Ia with similar $\dm15$ are more similar.  However, there is still
a large spread which is most likely due to errors in dereddening or
galaxy subtraction.
\label{fig:flux_X2.templcomp}}
\end{figure}

\begin{figure}[t]
\epsscale{0.9}
\ifsubmode
\plotone{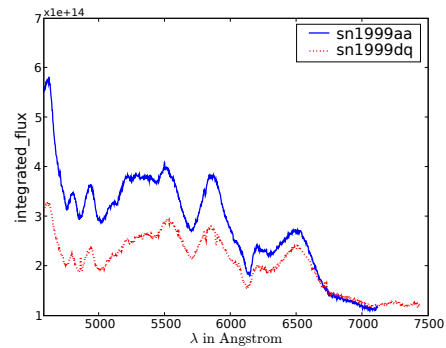}
\fi
\caption[]{
This figure shows the spectra of SN~1999dq (solid blue line) and
SN~1999aa (dotted red line). Note that the spectra of SN~1999aa
and SN~1999dq have almost identical spectral features but that of SN~1999aa is warped
toward the blue.  This difference is most likely not real but an
artifact of dereddening or background subtraction.
\label{fig:specexample_99aa_99dq}}
\end{figure}

We compare each SN Ia template spectrum with all others by performing a
$\chi^2$-minimization fit with a normalization factor as the only free
parameter. We use the variance spectrum as $1-\sigma$ errors for the
observed spectrum.  Figure~\ref{fig:flux_X2.templcomp} shows the
$\chi^2$ versus their difference in $\dm15$: $\Delta m_{15}(i)
- \Delta m_{15}(j)$ for all SN Ia with grade $A$ or $B$. There is a
trend that for smaller differences in $\dm15$ the $\chi^2$ is smaller,
indicating that indeed SN Ia with similar $\dm15$ have similar spectra.
However, there is still a big spread. This is most likely due to
errors in dereddening or galaxy subtraction  introducing systematic
errors. Figure\ref{fig:specexample_99aa_99dq} shows the spectra of
SN~1999dq and SN~1999aa.  The spectrum of SN~1999aa is very similar to
the spectrum of SN~1999dq, with the exception that the SN~1999aa
spectrum is warped toward the blue. This difference is most likely not
real but an artifact of dereddening or background subtraction.

In order to avoid these problems, one has to do a very careful
reduction and analysis to obtain the template SNe spectra. 
\begin{itemize}
\item Interpolate the observed $B$ and $V$ light curves
(no reddening corrections).
\item Warp the spectra to match the observed
$B-V$ color at a given epoch.
\item Calculate the K-correction from the warped
spectrum.
\item Apply calculated K-corrections and deredden the $B$ and $V$ light
curves to get the intrinsic $B-V$ color.
\item Finally correct the spectrum by
reddening and warping it to match the intrinsic color calculated from the
light curve.
\end{itemize}
Similar techniques are being developed for SN~Ia lightcurve fitting to
get distances to SNe~Ia. We are currently implementing these
techniques and we will discuss their implementation in \cite{Rest08}.

\subsubsection{Cross-Correlation of Spectra with \snid}
\label{sec:snid}

One alternative to the $\chi^2$-minimization approach is the
cross-correlation technique. In this paper, we use an implementation
of the correlation techniques of \citet{Tonry79}, the SuperNova
IDentification code (\snid;
\citealt{Blondin06,Blondin07}). In \snid, the
input and template spectra are binned on a common logarithmic
wavelength scale, such that a redshift $(1+z)$ corresponds to a
uniform linear shift in $\log \lambda$. The spectra are then
``flattened'' through division by a {\it pseudo} continuum, such that
the correlation only relies on the {\it relative} shape and strength
of spectral features, and is therefore insensitive to spectral color
information (including reddening uncertainties and flux
mis-calibrations). The pseudo-continuum is fitted as a 13-point cubic
spline evenly spaced in log wavelength between 2500\AA and 10000\AA.
We refer the reader to \cite{Blondin07} where the \snid\, algorithm is
described in full detail.  Finally, the spectra are smoothed by
applying a bandpass filter to remove low-frequency residuals
(wavelength scale $\gtrsim 300\rm{\AA}$) left over from the {\it
pseudo}-continuum division and high-frequency noise (wavelength scale
$\lesssim 50\rm{\AA}$) components.  The main motivation for applying
such a filter lies in the physical nature of supernova spectra, which
are dominated by broad spectral lines ($\sim 100-150\rm{\AA}$) due to
the large expansion velocities of the ejecta ($\sim
10000\rm{\kms}$). A more detailed explanation of the spectrum
pre-processing and cross-correlation in \snid\, is given by
\citet{Blondin07}. 

The input spectrum is correlated in turn with each template
spectrum. The redshift, $z$, is usually a free parameter in \snid\,
(indeed, this code was developed to determine the redshift of high-$z$
SN~Ia spectra;
\citealt{Matheson05,Miknaitis07}), but can be fixed
when the redshift is known, as is the case here for the LMC
($z\approx0.001$). The quality of a correlation is determined by the
$rlap$ quality parameter, which is the product of the Tonry \& Davis
correlation height-noise ratio ($r$) and the spectrum overlap
parameter ($lap$). $r$ is defined as the ratio of the height of the
heighest peak in the correlation function to the root-mean-square of
its antisymmetric component, while $lap$ is a measure of the overlap
in restframe wavelength space between the input and template spectra
\citep[see][]{Blondin07}. For an input spectrum with the
restframe wavelength range $[\lambda_0,\lambda_1]$, $lap$ is in the
range $0 \le lap \le \ln(\lambda_1/\lambda_0)$. In what follows, a
``good'' correlation corresponds to $rlap \geq 5$ with $lap \geq 0.4$
\citep{Matheson05,Miknaitis07,Blondin06,Blondin07}. Note that these limits are
not derived from the dataset in this paper, but from comparing single
epoch spectra from low and high-z SNe.

\begin{figure}[t]
\epsscale{0.9}
\ifsubmode
\plotone{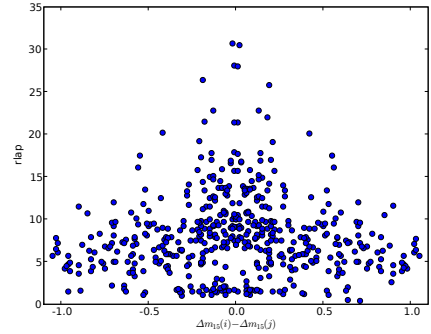}
\fi
\caption[]{This figure shows $rlap$ versus their
difference in $\dm15$: $\Delta m_{15}(i) - \Delta m_{15}(j)$ for all
SN Ia in category $A$ and $B$. There is a clear trend toward stronger
correlation for spectra with similar $\dm15$.
\label{fig:rlap.templcomp}}
\end{figure}

We correlate each pair of SN~Ia spectra using the \snid~terchniques in
a similar fashion what we have done with the $\chi^2$-minimization.
Figure~\ref{fig:rlap.templcomp} shows the $rlap$ values versus
difference in $\dm15$ of the \snid\ correlated Ia template spectra.
There is a clear trend toward stronger correlation for spectra with
similar $\dm15$.  Note that most of the pairs with $rlap \ge 10$ have
a difference in $\dm15$ of less than 0.25. 

\begin{figure}[t]
\epsscale{0.9}
\ifsubmode
\plotone{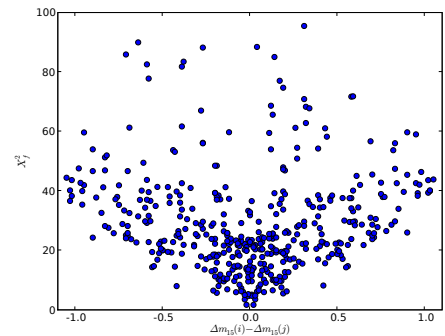}
\fi
\caption[]{This figure shows  $\chi_f^2$ versus their
difference in $\dm15$: $\Delta m_{15}(i) - \Delta m_{15}(j)$ for all
SN Ia in category $A$ and $B$. The correlation between the
goodness of fit indicated by $\chi_f^2$ and $\dm15$ is significantly
better than for $\chi^2$.
\label{fig:flat_X2.templcomp}}
\end{figure}

The \snid technique ``flattens'' the spectra before correlating them
(see Section~\ref{sec:snid}).  The flattened spectra produced
by \snid\ have the additional advantage that the $\chi^2$ is more
robust against errors in dereddening or background subtraction.  We
calculate $\chi_f^2$ for all SN Ia pairs of flattened template spectra
and plot it versus the difference in $\dm15$ in
Figure~\ref{fig:flat_X2.templcomp}.  The correlation between the
goodness of fit indicated by $\chi_f^2$ and $\dm15$ is significantly
better than for $\chi^2$.

\begin{figure}[t]
\epsscale{0.9}
\ifsubmode
\plotone{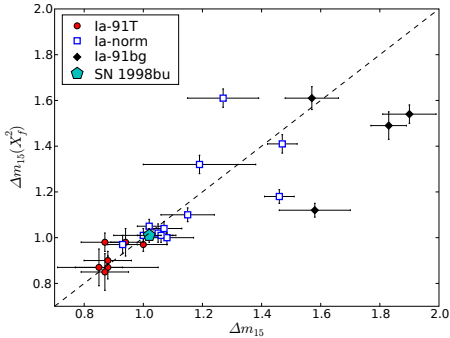}
\fi
\caption[]{This figure shows  $\dm15$
determined by the lightcurve shape versus the $\dm15$ estimated using
the $\chi_f^2$, denoted as $\dm15(\chi_f^2)$. The SN~Ia subtypes
``Ia-91T'', ``Ia-norm'', and ``Ia-91bg'' are indicated with filled red
circles, open blue squares, and filled black diamonds,
respectively. The cyan pentagon inidicates SN~1998bu. For SNe Ia with
small $\dm15 \le 1.1$, the agreement between true and estimated
$\dm15(\chi_f^2)$ is excellent.  However, for SNe Ia with $\dm15 >
1.1$, there is a significant spread.
\label{fig:dm15_X2_dm15}}
\end{figure}

\begin{figure}[t]
\epsscale{0.9}
\ifsubmode
\plotone{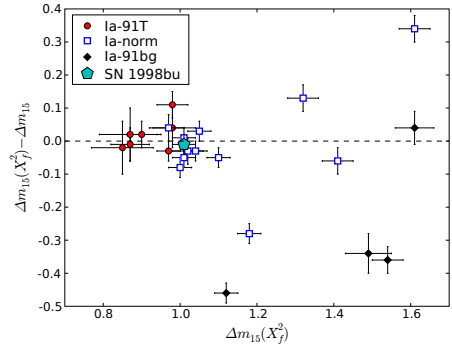}
\fi
\caption[]{This figure shows the difference between the true
$\dm15$ and the estimated $\dm15(\chi_f^2)$ versus $\dm15(\chi_f^2)$.
 The SN~Ia subtypes ``Ia-91T'', ``Ia-norm'', and ``Ia-91bg'' are
 indicated with filled red circles, open blue squares, and filled
 black diamonds, respectively. The cyan pentagon inidicates SN~1998bu.
 For $\dm15(\chi_f^2) \le 1.1$, the standard deviation of the
 estimated $\dm15(\chi_f^2)$ compared to the true $\dm15$ is 0.05
 magnitudes.
\label{fig:ddm15_dm15_X2}}
\end{figure}

We use the template spectra to test if $\chi_f^2$ can be used to
determine the $\dm15$: For a given SN Ia template spectra, we
calculate the $\chi_f^2$ with respect to all other SN Ia templates. We
find the template SNe with the smallest $\chi_f^2$ (noted
$\chi^2_{f,min}$), and estimate the $\dm15$ by computing the
error-weighted mean $\dm15$ for the three templates with the smallest
$\chi_f^2 \le 2 \times \chi^2_{f,min}$.
Figure~\ref{fig:dm15_X2_dm15} shows
the $\dm15$ determined by the lightcurve shape versus the $\dm15$
estimated using the $\chi_f^2$ (denoted as $\dm15(\chi_f^2)$. For
SNe Ia with small $\dm15 \le 1.1$, the agreement between true and
estimated $\dm15(\chi_f^2)$ is excellent. 
However, for SNe Ia with $\dm15 >
1.1$, there is a significant spread.  
The reason is that the sample of SN~Ia with $\dm15 < 1.1$
comprises both normal objects and overluminous ones with spectra
similar to SN~1991T or SN~1999aa \citep{Jeffery92,Jha06}. These latter objects have
spectra that show large deviations from normal SNe~Ia, especially
around maximum light (where the impact on the light-echo spectrum is
the greatest). At intermediate $\dm15$\ ($1.2 \lesssim \dm15
\lesssim 1.6$), however, SN~Ia spectra are similar, and our approach
does not enable a clear determination. The two points with large
$\dm15 > 1.8$ are both subluminous, 1991bg-like SNe~Ia \citep{Garnavich04,Jha06}. The
spectra show significant deviations from normal and overluminous
SNe~Ia around maximum light, yet the $\dm15$ values are
systematically underestimated. The disagreement is a simple
consequence of the lack of 1991bg-like SN~Ia in our set of spectral
templates. With only two such templates with $\dm15 > 1.6$ (see
Table~\ref{tab:snoverviewIa}, the error-weighted mean $\dm15$ of the
three best-matching templates will systematically bias the $\dm15$
determination to lower values.  In Figure~\ref{fig:ddm15_dm15_X2} we
show the difference between the true $\dm15$ and the estimated
$\dm15(\chi_f^2)$ versus $\dm15(\chi_f^2)$. For $\dm15(\chi_f^2) \le
1.1$, the standard deviation of the estimated $\dm15(\chi_f^2)$
compared to the true $\dm15$ is 0.05 magnitudes. The calculated
uncertainties of $\dm15(\chi_f^2)$ are slightly underestimated by a
factor of 1.3. Therefore we consider $\dm15(\chi_f^2)$ as a very good
estimate of the true $\dm15$ for $\dm15(\chi_f^2) \le 1.1$ with
uncertainties smaller than 0.1 magnitudes.  If the fitted $\dm15 >
1.1$, then it can only be said that it is unlikely to be a 1991T-like
SN~Ia (1999cl is the only 91T-like SN~Ia in our sample with $\dm15 >
1.1$; see Table~\ref{tab:snoverviewIa}).  More importantly,
Figure~\ref{fig:ddm15_dm15_X2} also shows that we are able to
accurately determine the $\dm15$ for SNe~Ia with $\dm15 < 1.1$. Since
all 1991T-like SN~Ia fulfill this condition, we are in principle able
to not only determine the $\dm15$ but also the SN~Ia subtype for these
objects. We will see in the next section that the light-echo spectrum
presented here is most probably a 1991T-like SN~Ia with $dm15 < 0.9$.

\subsection{Single Scattering Approximation}
\label{sec:scatter}

Several authors have addressed the single scattering approximation for
light echoes
\citep[e.g.]{Couderc39,Chevalier86,Emmering89,Sugerman03,Patat05}.
Following the derivation of \cite{Sugerman03}, the surface brightness
$B_{SC}$ of scattered light with wavelength $\lambda$ scattering at an
angle $\theta$ off dust at position $\rb$ and thickness $\Delta z$ is
\begin{equation}
B_{SC}(\lambda,\theta,\rb,\Delta z) = F(\lambda) n_H(\rb) G(\rb,\Delta z) S(\lambda,\theta)
\end{equation}
Where $F(\lambda)$ is the time-integrated spectra of the SN, $n_H(\rb)$
is the number density of hydrogen nuclei, $G(\rb,\Delta z)$ is a
geometrical factor depending on the geometry between the observer, SN,
and dust, and the integrated scattering function $S(\lambda,\theta)$
which is described in more detail in Section~\ref{sec:dustproperties}.

Since for this paper we are only interested in the relative fluxes, we
can drop all terms that are not dependent on the wavelength, add
reddening by the LMC and MWG, and the modelled, observed spectrum
$F_{scat}$ has the form
\begin{equation}
F_{scat}(\lambda) = C_{norm} C_{ext}^{MWG}(\lambda) C_{ext}^{LMC}(\lambda) F(\lambda) S(\lambda,\theta) \label{eq:fitfcn}
\end{equation}
where $C_{ext}^{MWG}$, and $C_{ext}^{LMC}$ are the reddening by the
MWG and the LMC, respectively, as discussed in
Section~\ref{sec:extinction}.

\subsubsection{Dust Properties}
\label{sec:dustproperties}
In order to get the total integrated scattering function $S(\lambda,\theta)$, we add up the 
integrated  scattering function for each individual dust type:
\begin{equation}
S(\lambda,\theta) = \sum_X S_X(\lambda,\theta)
\end{equation}
Following \cite{WD01}, $X$ can be $s$, $cn$, and $ci$ for silicon dust
grains, carbonaceous dust grains with neutral PAH component, and
carbonaceous dust grains with an ionized PAH component, respectively,
where PAH stands for Polycyclic Aromatic Hydrocarbon. For a given dust
type $X$, the integrated scattering function $S_X$ is
\begin{equation}
S_X(\lambda,\theta) = \int Q_{SC,X}(\lambda,a)\sigma_g\Phi_X(\theta,\lambda,a)f_X(a)da
\end{equation}
where $Q_{SC,X}$ is the grain scattering efficiency, $\sigma_g=\pi a^2$ is the grain cross-section, $f_X(a)$
is the grain size distribution discussed in Section~\ref{sec:dustgrains}, and 
$\Phi_X(\theta,\lambda,a)$ is the \cite{Henyey41} phase function
\begin{equation}
\Phi_X(\lambda,\theta) = \frac{1-g_X^2(\lambda,a)}{(1+g_X^2(\lambda,a)-2g_X(\lambda,a)\cos{\theta})^{3/2}}
\end{equation}
with $g(\lambda,a)$ the degree of forward scattering for a given
grain.  We integrate $S_X(\lambda,\theta)$ for the individual grain
types by using the extended Simpson's method \citep{Press92}. We
interpolate the values for $Q_{SC,X}$ and $g_X(\lambda,a)$ using the
tables provided by
B. T. Draine\footnote{http://www.astro.princeton.edu/$\sim$draine/dust/dust.diel.html}
\citep{Draine84,Laor93,WD01,Li01}.

\subsubsection{Dust Grain Size Distribution}
\label{sec:dustgrains}

We use the models defined by \cite{WD01} which consist of mixture of
carbonaceous grains and amorphous silicate grains. Carbonaceous grains
are PAH-like when small ($a \leq 10^{-3} \micron$), and graphite-like when large
($a>10^{-3} \micron$)\citep{Li01}. The dust
grain size distribution $f(a)$ is written as
\begin{equation}
f(a) \equiv \frac{1}{n_H}  \frac{dn_{gr}}{da}
\end{equation}
with $n_{gr}(a)$ is the number density of grains with size $\le a$ and
$n_H$ is the number density of H nuclei in both atoms and
molecules. \cite{WD01} derives the size distributions for different
line of sights toward the LMC. We adopt the parameters of their
``LMC avg'' model and $b_C=2\times 10^{-5}$ \citep[Table 3,][]{WD01}.  We can
then calculate the size distributions for carbonaceous dust
$f_{ci}(a)=C_{ion} f(a)$ and $f_{cn}(a)=(1-C_{ion}) f(a)$ using
equations 2, 4, and 6 of \cite{WD01}.  The PAH/graphitic grains are
assumed to be 50\% neutral and 50\% ionized \citep{Li01}, thus
$C_{ion}=0.5$. For amorphous silicate dust $f_{s}$ we use equations 5
and 6.

\subsubsection{Extinction}
\label{sec:extinction}
The extinction can be expressed as
\begin{equation}
\log_{10} C_{ext}(\lambda) = -0.4 \langle A(\lambda)/A(V) \rangle R_V E(B-V)
\end{equation}
For the MWG, we calculate the extinction $C_{ext}^{MWG}(\lambda)$
setting $R_V=3.1$ and $E_{MWG}(B-V) = 0.07$, and by calculating
$\langle A(\lambda)/A(V) \rangle$ using Equations~1-3 of
\cite{Cardelli89}.

The average internal extinction of the LMC is $E(B-V) = 0.1$
\citep{Bessel91}, but different populations give different
results. \cite{Zaritsky99} finds a mean $E(B-V)=0.06$ from red clump
giants and $E(B-V)=0.14$ from OB types. They attribute this dependence
to an age-dependent scale height: OB stars having a smaller scale
height lie predominantly in the dusty disk. We use $E_{LMC}(B-V) =
0.05$, half the average internal extinction value, since the most likely 
position of the SN is somewhere halfway through the LMC. Since the
light echoes are not in the Superbubble of the LMC, we use the average
$R_{V,LMC}=3.41$ value of the ``LMC Average Sample'' in Table 2,
\cite{Gordon03}. Even though the extinction curves in the LMC and SMC
have similarities to the MWG extinction curves, there are significant
differences. Thus we calculate $\langle A(\lambda)/A(V) \rangle$ for
the LMC using Equation~5 in \cite{Gordon03} and the values in the
``Average'' row of the section ``LMC Average Sample'' in Table~3 in
\cite{Gordon03}.

\subsection{Testing the method with SN~1998bu}
\label{sec:98bu}

Several 100 days after the explosion, the light curve of the type Ia
supernova SN~1998bu suddenly flattened. At the same time, the spectrum
changed from the typical nebular emission to a blue continuum with
broad absorption and emission features reminiscent of the SN spectrum
at early phases \citep{Cappellaro01}. This was explained by the
emergence of a light echo from a foreground dust cloud. A similar case
is SN~1991T, but its light echo spectra are of significantly less
signal-to-noise.

We use SN~1998bu as a test cases to see if we can determine the
$\dm15$ of the SNe~Ia from their light echo spectrum. We correlate the
template spectra with the light echo spectrum and estimate with our
method the $\dm15(\chi_f^2)$ of SN~1998bu. We can then compare how
close this estimated $\dm15$ is to the true, lightcurve measured
$\dm15=1.02 \pm 0.04$. This is the ultimate test if the method works
on a real world example.

For the light echo of SN~1998bu
we assume that the reflecting dust is $z=70 pc$ in front of the
supernova, the host galaxy extinction is $A_V^{host}=0.86$, and
$R_V=3.1$ \citep{Cappellaro01}.  
Using these values, we can calculate time-integrated, dust scattered,
and reddened template spectra by applying equation~\ref{eq:fitfcn} for
28 SNe~Ia, which we denote 
template spectra in what follows.  As described in the previous
Section~\ref{sec:snid}, we calculate the $rlap$ and $\chi_f^2$ values using \snid\ 
for the observed light echo spectra with the spectra templates.

\begin{figure}[t]
\epsscale{0.9}
\ifsubmode
\plotone{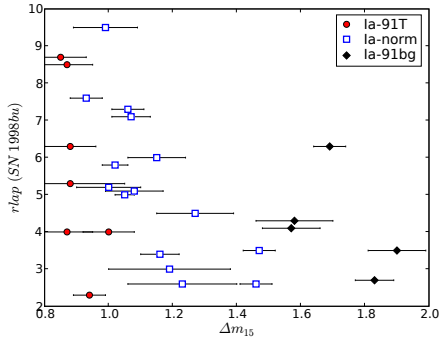}
\fi
\caption[]{Correlation parameter $rlap$ of the light echo of SN~1998bu determined with \snid\ versus \dm15\, of
the time-integrated, dust-scattered, reddened, and flattened SN~Ia
spectra. The SN~Ia subtypes ``Ia-91T'', ``Ia-norm'', and ``Ia-91bg''
are indicated with filled red circles, open blue squares, and filled
black diamonds, respectively.
\label{fig:98bu_rlap_dm15}}
\end{figure}

\begin{figure}[t]
\epsscale{0.9}
\ifsubmode
\plotone{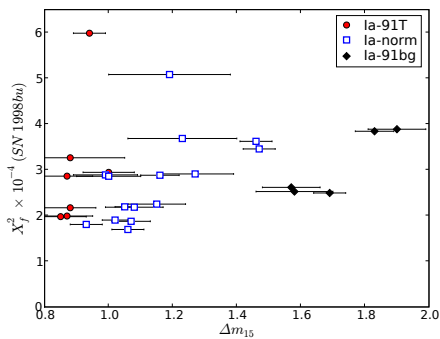}
\fi
\caption[]{$\chi_f^2$ versus \dm15 for the fit of the time-integrated,
dust-scattered, reddened, and {\it flattened} SN~Ia spectra to the
observed light echo spectrum associated with SN~1998bu.  The SN~Ia
subtypes ``Ia-91T'', ``Ia-norm'', and ``Ia-91bg'' are indicated with
filled red circles, open blue squares, and filled black diamonds,
respectively.
\label{fig:98bu_flatX2_dm15}}
\end{figure}

Figure~\ref{fig:98bu_rlap_dm15} shows the $rlap$ values of the
SN~1998bu light echo and the template spectra for the different SN~Ia
subtypes
In general, 91T-like SNe~Ia (filled red circles) have a lower than
normal
\dm15 and are overluminous, slow decliners, whereas 91bg-like SNe~Ia
(filled black diamonds) have a large \dm15 and are underluminous, fast
decliners. Besides one outlier at $\dm15=1.7$, all other $rlap \ge 5$
values are for SN Ia with $\dm15<1.2$. This is in very good agreement
with the $\dm15 = 1.02 \pm 0.04$ of SN~1998bu determined from its
lightcurves (Note that the limit $rlap \ge 5$ is not derived from the
dataset in this paper, but from comparing single epoch spectra from
low and high-z SNe). Similarly, the best $\chi_f^2$ values are all for
spectra templates with $\dm15<1.2$ (see
Figure~\ref{fig:98bu_flatX2_dm15}). We apply our method to determine
the $\dm15$ described in Section~\ref{sec:snid} using the $\chi_f^2$
and estimate the $\dm15(\chi_f^2)=1.01 \pm 0.03$ for SN~1998bu. This
is within the errors to the $\dm15$ determined with the
lightcurves. Figure~\ref{fig:dm15_X2_dm15}~and~\ref{fig:ddm15_dm15_X2}
show the $\dm15(\chi_f^2)$ of SN~1998bu (red circle).  We find that
the \snid\ correlation technique provides more than sufficient
discrimination between input templates for $\dm15 \le 1.1$


\section{Discussion}
\label{sec:discussion}

We have obtained a spectrum of the light echo located at
RA=05:13:03.77 and DEC=-67:29:04.91 at epoch J2000.  The associated
SNR 0509-675 is at RA=05:09:31.922 and DEC=-67:31:17.12 at epoch J2000.  The
angular distance between the light echo and the SNR is 0.340 degrees.
Using the SNR age of 400 years determined using light echo apparent
motion
\citep{Rest05b}, we can determine the line-of-sight distance between the
dust sheet and the SNR. We find this distance to be $z=300$ pc, which
we use to calculate the scattering angle $\theta$. Then we can
calculate the time-integrated, dust-scattered, and reddened template
spectra by applying equation~\ref{eq:fitfcn} for 28 SNe~Ia and 6
SNe~Ib/c. We use \snid\ to calculate the $rlap$ and $\chi_f^2$ using
both the SNe Ia and SNe~Ib/c templates (see Column~(9)~and~(10) in
Table~\ref{tab:snoverviewIa} and Column~(8)~and~(9) in
Table~\ref{tab:snoverviewIbc}).

\begin{figure}[t]
\epsscale{0.9}
\ifsubmode
\plotone{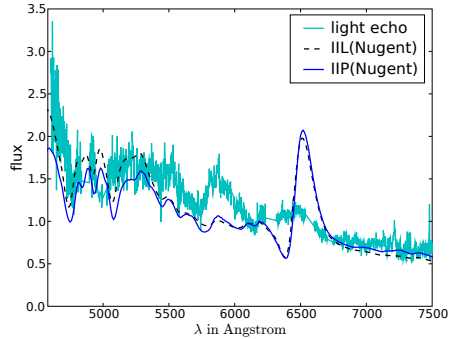}
\fi
\caption[]{
Time-integrated, dust-scattered, and reddened SN II spectra fitted to
the observed light echo spectrum (cyan line) associated with SNR
0509-675.  The chip gaps (4838\AA-4894\AA, 6235\AA-6315\AA), and
various LMC background emission lines ([O III], H$\alpha$, [N II], [S
II]) are removed from the observed spectrum.  The features do not
correlate, and we conclude that the observed light echo spectrum is
not a SN II spectrum.
\label{fig:SNIIfit}}
\end{figure}

The spectrum of the light echo associated with SNR 0509-675 shows
broad emission and absorption features consistent with SN spectra. The
question is, what kind of SN is it?  Figure~\ref{fig:SNIIfit} shows
the fit of two SN II template spectra (Type IIP and IIL) to the echo
spectrum.  Both spectra are created by using a spectral library by
Nugent\footnote{\texttt{http://supernova.lbl.gov/$\sim$nugent/nugent\_templates.html}}
\citep{Gilliland99}.
The IIP spectra template is based mostly on the models seen in
\cite{Baron04},  and its lightcurves are based on \cite{Cappellaro97}.
There is no need for any sophisticated correlation or fitting
technique to conclude that the observed light echo spectrum is not
that of a Type II supernova.

\begin{figure}[t]
\epsscale{0.9}
\ifsubmode
\plotone{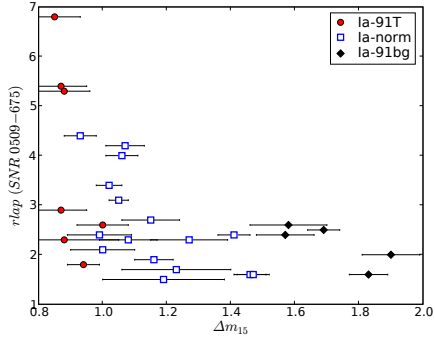}
\fi
\caption[]{
Correlation parameter $rlap$ determined with \snid\ versus \dm15\, of
the time-integrated, dust-scattered, reddened, and flattened SN~Ia
spectra. The SN~Ia subtypes ``Ia-91T'', ``Ia-norm'', and ``Ia-91bg'' are indicated
with filled red circles, open blue squares, and filled black diamonds,
respectively. An $rlap$ value bigger than 5 is considered a strong
correlation.  There is a clear trend toward template spectra with
small \dm15 correlating stronger with the observed light echo spectrum
than the ones with large \dm15. Only three templates, all with
$\dm15<0.9$, have a strong correlation with the observed light echo
spectrum ($rlap>5.0$).
\label{fig:rlap_dm15}}
\end{figure}

\begin{figure}[t]
\epsscale{0.9}
\ifsubmode
\plotone{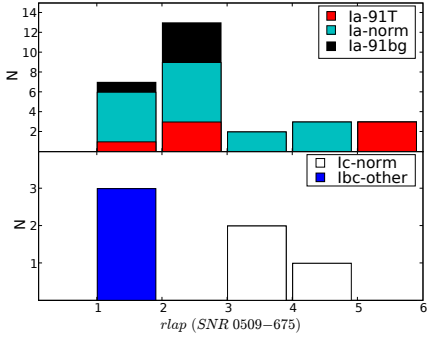}
\fi
\caption[]{
Histogram of $rlap$ value determined with \snid\ for the different SN~Ia
subtypes (``Ia-91T'', ``Ia-norm'', and ``Ia-91bg'', upper panel), normal SN~Ic type
(``Ic-norm'', lower panel), and SN~Ib/c type other than normal Ic (``Ibc-other'',
lower panel). An $rlap$ value bigger than 5 indicates a good
correlation between the two spectra. Note that the 91T-like SNe have
the strongest correlation with the light echo spectrum. All SNe~Ib/c
have a significantly smaller $rlap$ value than three of the 91T-like
SNe~Ia, and no SNe~Ib/c has an $rlap$ value bigger than 5, which is
the cutoff value for a good correlation.
\label{fig:rlap_histo}}
\end{figure}

Following the procedures described in
Section~\ref{sec:scatter}~and~\ref{sec:timeintspec}, we have created
the SN template spectra for 28 SNe~Ia and 6 SNe~Ib/c, and correlate
these template spectra with the light echo spectrum using the \snid\
correlation technique described in Section~\ref{sec:snid}. For the
purpose of this paper, which is to identify the spectral (sub)type of
the SN explosion that created SNR 0509-675, the \snid\ correlation
technique described in Section~\ref{sec:snid} provides more than
sufficient discrimination between input templates.  This technique
``flattens'' both spectra, and then correlates the main features of
the spectra. The strength of the correlation is reflected in the
parameter $rlap$, with values of $rlap \ge 5.0$ indicating a strong
correlation.  Figure~\ref{fig:rlap_dm15} shows the $rlap$ values
versus \dm15 of the \snid\ correlated Ia template spectra.  The SN~Ia
subtypes ``Ia-91T'', ``Ia-norm'', and ``Ia-91bg'' are indicated with
filled red circles, open blue squares, and filled black diamonds,
respectively.  There is an improved match with template spectra with
smaller \dm15 correlating more strongly with the observed light echo
spectrum than the ones with large \dm15. Only three templates, all
with $\dm15<0.9$, have a strong correlation with the observed light
echo spectrum. Figure~\ref{fig:rlap_histo} shows a histogram of the
$rlap$ values for the different SN~Ia as well as Ib/c subtypes.  Note
that the normal SNe~Ic (SN~1994I, SN~2004aw, SN~2005mf), denoted as
``Ic-norm'', show stronger correlations than the other types of Ib/c
(the broad-line SN~Ic SN~1997ef, the peculiar SN~Ib SN~2005bf, and the
normal SN~Ib SN 2005hg), denoted as ``Ibc-other'' in the lower panel
of Figure~\ref{fig:rlap_histo} (see also
Table~\ref{tab:snoverviewIbc}).  All SNe~Ib/c have a significantly
smaller $rlap$ value than three of the 91T-like SN~Ia SN, and no
SN~Ib/c has an $rlap$ value bigger than 5, which is the cutoff value
for a good correlation.

\begin{figure}[t]
\epsscale{0.9}
\ifsubmode
\plotone{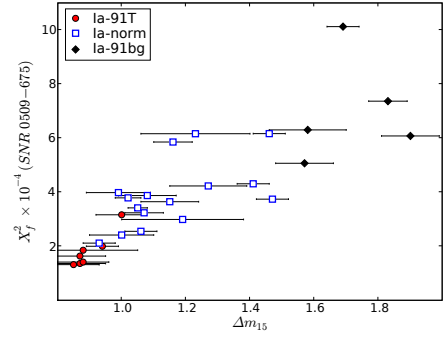}
\fi
\caption[]{
$\chi_f^2$ versus \dm15 for the fit of the time-integrated,
dust-scattered, reddened, and {\it flattened} SN~Ia spectra to the
observed light echo spectrum associated with SNR 0509-675. The SN~Ia
subtypes ``Ia-91T'', ``Ia-norm'', and ``Ia-91bg'' are indicated with filled red
circles, open blue squares, and filled black diamonds, respectively.
The 5 SNe with the smallest $\chi_f^2$ are 91T-like SNe Ia and have
the smallest \dm15. Note that the goodness of fit correlates better
with \dm15 than subtype. The 91T-like SNe with comparably large
\dm15 have a larger $\chi_f^2$ than the normal Ia with small \dm15.
\label{fig:flatX2_dm15}}
\end{figure}

\begin{figure}[t]
\epsscale{0.9}
\ifsubmode
\plotone{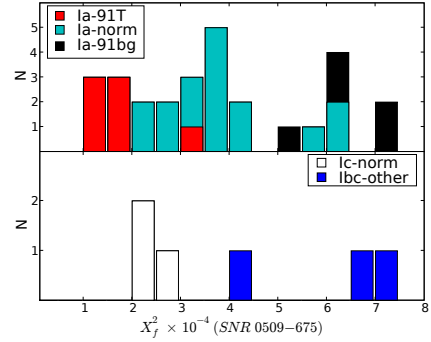}
\fi
\caption[]{
The upper panel shows the histogram of $\chi_f^2$ for the fit of the
time-integrated, dust-scattered, reddened, and {\it flattened} SN~Ia
spectra fit to the {\it flattened} observed light echo spectrum
associated with SNR 0509-675. Note that nearly all 91T-like SNe Ia
have a better $\chi_f^2$ than any SNe of the other subtypes. The lower
panel shows the same histogram of $\chi_f^2$ for the SNe~Ib/c.  Note
that the normal SNe~Ic have a decent $\chi_f^2$, but all 91T-like SNe
with $\dm15<0.9$ have a better $\chi_f^2$.
\label{fig:flatX2_histo}}
\end{figure}

The flattened spectra produced by \snid\ have the additional advantage
that the $\chi^2$ is more robust against errors in dereddening or
background subtraction.  We calculate $\chi_f^2$ of the flattened
template spectra with respect to the flattened observed light echo
spectra for all templates (see Column~(9) in
Table~\ref{tab:snoverviewIa} and Column~(8) in Table
\ref{tab:snoverviewIbc}).  Figure~\ref{fig:flatX2_dm15} shows
$\chi_f^2$ versus \dm15 for the different subtypes of SN~Ia. The SN~Ia
subtypes ``Ia-91T'', ``Ia-norm'', and ``Ia-91bg'' are indicated with
filled red circles, open blue squares, and filled black diamonds,
respectively.  The correlation between $\chi_f^2$ and \dm15 is, as
expected, excellent: The 5 SNe with the smallest $\chi_f^2$ also have
the smallest \dm15. This confirms that $rlap$ and $\chi^2$ are equally
suitable measures of fit when the low spatial frequency feautures are
removed.  Figure~\ref{fig:flatX2_histo} shows the histograms of
$\chi_f^2$ for the different SN~Ia subtypes (upper panel) and SN~Ib/c
subtypes (lower panel). Similar to the $rlap$ histograms, the normal
SNe~Ic have a decent $\chi_f^2$, but all 91T-like SNe with $\dm15<0.9$
have a better $\chi_f^2$. We apply our method to determine the $\dm15$
of the light echo spectra using the $\chi_f^2$, and we find that
$\dm15(\chi_f^2)= 0.87 \pm 0.05$.

\begin{figure}[t]
\epsscale{0.9}
\ifsubmode
\plotone{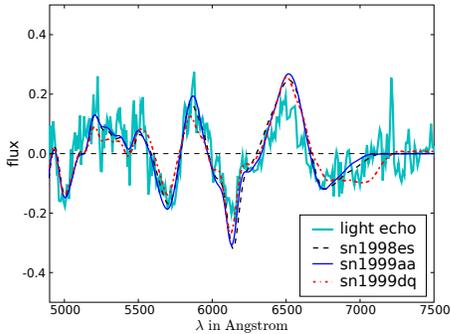}
\fi
\caption[]{
The three time-integrated, dust-scattered, reddened, and {\it
flattened} SN~Ia spectra with the best $rlap$ values (i.e. the best
correlation with the observed light echo spectrum), overplotted on the
{\it flattened} observed light echo spectrum (solid cyan line). Note
that all three template spectra are 91T-like SNe~Ia with $\dm15<0.9$.
\label{fig:91T.flattened}}
\end{figure}

\begin{figure}[t]
\epsscale{0.9}
\ifsubmode
\plotone{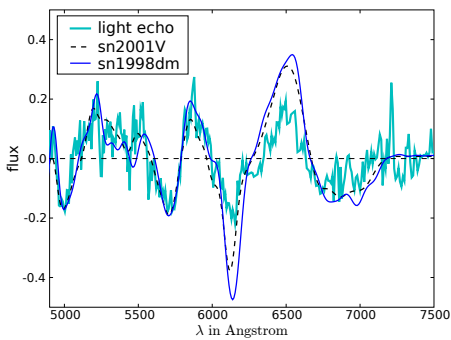}
\fi
\caption[]{
The two time-integrated, dust-scattered, reddened, and {\it flattened}
spectra of normal SNe~Ia with the best rlap values, overplotted on the
{\it flattened} observed light echo spectrum (solid cyan line). Note
that these two template spectra are also the two normal SNe~Ia with
the smallest
\dm15.
\label{fig:Ia-norm.flattened}}
\end{figure}

\begin{figure}[t]
\epsscale{0.9}
\ifsubmode
\plotone{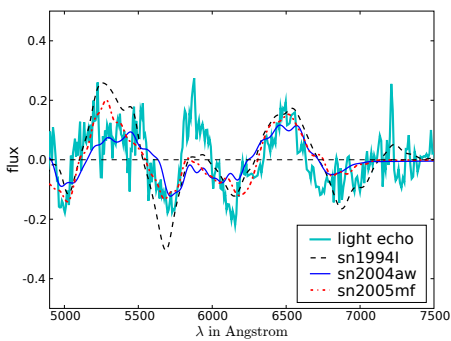}
\fi
\caption[]{
The three time-integrated, dust-scattered, reddened, and {\it
flattened} SN~Ib/c spectra with the best $rlap$ values, overplotted on
the {\it flattened} observed light echo spectrum (solid cyan
line). Note that all three template spectra are normal SNe~Ic.
\label{fig:Ic-norm.flattened}}
\end{figure}

\begin{figure}[t]
\epsscale{0.9}
\ifsubmode
\plotone{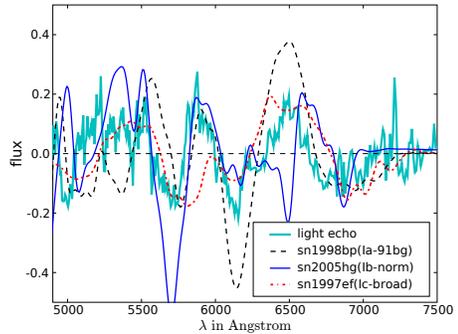}
\fi
\caption[]{
Three time-integrated, dust-scattered, reddened, and {\it flattened}
SN template spectra of different types having a bad correlation with
the observed light echo spectrum (cyan line). 
\label{fig:badcorr.flattened}}
\end{figure}

Figure~\ref{fig:91T.flattened} shows the three 91T-like SN~Ia template
spectra with the best $rlap$ values, overplotted on the observed light
echo spectrum. The template spectra have the same features than the
observed light echo spectrum, and that the agreement is very good. The
only significant disagreement is that the template spectra have a
slightly deeper absorption and stronger emission at $6100$\AA\ and
$6500$\AA, respectively.  The two ``normal'' SNe~Ia with the best
$rlap$ values are very similar and fit as well very good for wavelength smaller
than $5800$\AA, but the difference in strength of the two features at
$6100$\AA\ and $6500$\AA is more pronounced (see
Figure~\ref{fig:Ia-norm.flattened}).  Note also that these two
best-correlating normal SNe~Ia are also the ones with the smallest
\dm15 values. Figure~\ref{fig:Ic-norm.flattened} shows the
three SN~Ib/c template spectra with the best $rlap$ values.  Even
though there are similarities between the template spectra and the
observed spectrum, the relative strength of the spectral features is
not as consistent as for the SNe~Ia. A few examples of SN template
spectra with a poor correlation to the observed light echo spectrum are
shown in Figure~\ref{fig:badcorr.flattened}.

We conclude that the SN that created SNR 0509-675 is a 91T-like SN
with $\dm15<0.9$. Knowing the SN type (and moreover its subtype and
thus how energetic the explosion was) places stringent constraints on
the explosion mechanism and hence on the interpretation of X-ray
spectra of the remnant. Analysis of X-ray data of SNR 0509-675 by
\cite{Hughes95} classifies this SNR as a remnant of a SN~Ia. Recent
analysis of X-ray spectra by \cite{Badenes08} also supports the
classification as an overluminous, 91T-like SN Ia: models using
hydrodynamic calculations and nonequilibrium ionization simulations of
highly energetic SN Ia reproduce the X-ray spectrum with its line flux
ratios better than normal or subenergetic models
\citep{Badenes08}. This is the first time that the (sub)type of an
ancient SN has been determined by direct means by taking the spectrum
of the original event.

\section{Conclusions}
\label{sec:conclusions}

We have obtained a spectrum of a light echo associated with SNR
0509-675.  By comparing and correlating time-integrated,
dust-scattered, and reddened template spectra created from a spectral
library of nearby supernovae of all types to the light echo spectrum,
we find that overluminous, 91T-like SNe~Ia match the observed spectrum
best. We correlate the template spectra with the observed spectra with
\snid.  The correlation parameter $rlap$ is a measure of the strength 
of the correlation, and $rlap \ge 5$ indicates a strong
correlation. Only SN~Ia with $\dm15<0.9$ pass this cut. They
correspond to intrinsically overluminous SNe~Ia with a spectrum
resembling the prototypical SN~1991T \citep{Jeffery92}.  Similarly,
{\it all} 91T-like SN~Ia with $\dm15<0.9$ have a smaller $\chi^2$ than
any other SNe when the ``flattened'' (see Section~\ref{sec:snid}) SN
templates are fitted to the observed spectrum.  Normal SNe~Ic show
some similarities to the observed light echo spectrum. However, the
correlation is only weak ($rlap<5$) and their $\chi^2$ is larger than
the $\chi^2$ of the 91T-like SN~Ia. Thus we can exclude them as the
possible source event for SNR 0509-675.  This is the first time that
the (sub)type of a SN is conclusively and directly determined long
after the event happened. Light echoes provide an excellent
opportunity to connect the physics of the SN itself to its
remnant. Much can be learned about the physics of SNe and their impact
on the surrounding ISM from this direct comparison: Knowing the SN
type (and moreover its subtype and thus how energetic the explosion
was) places stringent constraints on the explosion mechanism and hence
on the interpretation of X-ray spectra of the remnant
\citep{Badenes08}. For the first time, models of SN explosions can now
be tested for their mutual conistency with the SNe explosion itself
and the observations of the SNR. We are currently working on expanding
the sample of SNR with light echo spectra: In the LMC alone there are
two more SNR with associated light echoes.  Our investigation suggests
also that subtyping of historical Milky Way supernovae, particularly
the more recent SN~1572 (Tycho), SN~1604 (Kepler) and Cas A events,
should be possible provided that suitable light echo features are
found and can be studied spectroscopically. Such a sample of SNRs with
known explosion spectra will place stringent constrains on SNe
explosion models and enhance our understanding of these events that
play such an important role in the production of heavy elements in our
universe.

\section{Acknowledgments}

Based on observations obtained for programs GS-2005B-Q-11 and
GS-2006B-Q-41 at which is operated by the Association of Universities
for Research in Astronomy, Inc., under a cooperative agreement with
the NSF on behalf of the Gemini partnership: the National Science
Foundation (United States), the Science and Technology Facilities
Council (United Kingdom), the National Research Council (Canada),
CONICYT (Chile), the Australian Research Council (Australia), CNP
(Brazil) and CONICEI (Argentina). The SuperMACHO survey was conducted
under the auspices of the NOAO Survey Program. The support of the
McDonnell Foundation, through a Centennial Fellowship awarded to
C. Stubbs, has been essential to the SuperMACHO survey. AR thanks the
Goldberg Fellowship Program for its support.  DW acknowledges the
support of a Discovery Grant from the Natural Sciences and Engineering
Research Council of Canada (NSERC).  AC is supported by FONDECYT grant
1051061. AG thanks the University of Washington Department of
Astronomy for facilities support.  DM, GP, LM and AC are supported by
Fondap Center for Astrophysics 15010003. The work of MH, KHC and SN
was performed under the auspices of the U.S. Department of Energy by
Lawrence Livermore National Laboratory in part under Contract
W-7405-Eng-48 and in part under Contract DE-AC52-07NA27344.  MWV is
supported by AST-0607485. GP acknowledges support by the Proyecto
FONDECYT 3070034. We thank Geoffrey C. Clayton for fruitful
discussions on how to implement the LMC internal reddening. This
research has made use of the CfA Supernova Archive, which is funded in
part by the National Science Foundation through grant AST 0606772.


\bibliographystyle{apj}
\bibliography{ms}



\clearpage
\begin{deluxetable}{llrrrrcrrrr}
\tabletypesize{\scriptsize} 
\tablecaption{
\label{tab:snoverviewIa}}
\tablehead{
\colhead{SN} & \colhead{Subtype} & \colhead{\dm15} & \colhead{N} & \colhead{$p_{min}$} & \colhead{$p_{max}$} & \colhead{$w_{max}$} & \colhead{Grade} & \colhead{$\chi^2_f*10^{-4}$} & \colhead{rlap} \\
\colhead{(1)} & \colhead{(2)} & \colhead{(3)} & \colhead{(4)} & \colhead{(5)} & \colhead{(6)} & \colhead{(7)} & \colhead{(8)} & \colhead{(9)} & \colhead{(10)} 
}
\startdata
  sn1991T&  Ia-91T&$0.94 \pm 0.05$&   18&     -15.50&      71.50&      0.2041&    A&                2.00& 1.8&       \\
  sn1992A& Ia-norm&$1.47 \pm 0.05$&   11&      -7.50&      25.50&      0.2087&    A&                3.74& 1.6&       \\
 sn1997do& Ia-norm&$0.99 \pm 0.10$&   12&     -12.05&      20.79&      0.2878&    B&                3.98& 2.4&       \\
  sn1998V& Ia-norm&$1.06 \pm 0.05$&    8&      -1.99&      41.99&      0.3026&    B&                2.55& 4.0&       \\
 sn1998ab&  Ia-91T&$0.88 \pm 0.17$&   10&      -8.13&      46.74&      0.4090&    C&                1.85& 2.3&       \\
 sn1998aq& Ia-norm&$1.05 \pm 0.03$&   14&      -1.24&      48.67&      0.2990&    B&                3.42& 3.1&       \\
 sn1998bp& Ia-91bg&$1.83 \pm 0.06$&   11&      -5.03&      27.85&      0.2189&    A&                7.37& 1.6&       \\
 sn1998bu& Ia-norm&$1.02 \pm 0.04$&   26&      -3.36&      43.66&      0.2159&    A&                3.79& 3.4&       \\
 sn1998dh& Ia-norm&$1.23 \pm 0.17$&   10&     -10.07&      44.82&      0.4386&    C&                6.16& 1.7&       \\
 sn1998dm& Ia-norm&$1.07 \pm 0.06$&   10&       2.91&      45.86&      0.4007&    C&                3.24& 4.2&       \\
 sn1998ec& Ia-norm&$1.08 \pm 0.09$&    6&      -3.98&      39.01&      0.2545&    B&                3.88& 2.3&       \\
 sn1998eg& Ia-norm&$1.15 \pm 0.09$&    6&      -0.35&      23.68&      0.3581&    C&                3.65& 2.7&       \\
 sn1998es&  Ia-91T&$0.87 \pm 0.08$&   20&     -12.28&      43.72&      0.1753&    A&                1.36& 5.4&       \\
 sn1999aa&  Ia-91T&$0.85 \pm 0.08$&   22&     -11.16&      47.64&      0.2044&    A&                1.32& 6.8&       \\
 sn1999ac&  Ia-91T&$1.00 \pm 0.08$&   16&      -5.97&      38.96&      0.1817&    A&                3.16& 2.6&       \\
 sn1999by& Ia-91bg&$1.90 \pm 0.09$&   13&      -5.36&      40.67&      0.1568&    A&                6.08& 2.0&       \\
 sn1999cc& Ia-norm&$1.46 \pm 0.05$&    7&      -4.14&      24.82&      0.2495&    A&                6.17& 1.6&       \\
 sn1999cl&  Ia-91T&$1.19 \pm 0.19$&   11&      -9.32&      36.67&      0.2792&    B&                2.99& 1.5&       \\
 sn1999dq&  Ia-91T&$0.88 \pm 0.08$&   21&     -13.02&      45.84&      0.1597&    A&                1.41& 5.3&       \\
 sn1999ej& Ia-norm&$1.41 \pm 0.05$&    5&      -2.23&      10.76&      0.4068&    C&                4.31& 2.4&       \\
 sn1999gd& Ia-norm&$1.16 \pm 0.06$&    5&       0.02&      33.89&      0.4036&    C&                5.86& 1.9&       \\
 sn1999gh& Ia-91bg&$1.69 \pm 0.05$&   12&       2.02&      38.88&      0.4359&    C&               10.12& 2.5&       \\
 sn1999gp&  Ia-91T&$0.87 \pm 0.08$&    8&      -6.29&      34.62&      0.1947&    A&                1.64& 2.9&       \\
 sn2000cf& Ia-norm&$1.27 \pm 0.12$&    6&      -0.16&      22.83&      0.3560&    C&                4.23& 2.3&       \\
 sn2000cn& Ia-91bg&$1.58 \pm 0.12$&    9&     -10.14&      26.77&      0.2764&    B&                6.30& 2.6&       \\
 sn2000dk& Ia-91bg&$1.57 \pm 0.09$&    6&      -5.11&      33.77&      0.2575&    B&                5.07& 2.4&       \\
 sn2000fa& Ia-norm&$1.00 \pm 0.10$&   13&     -12.03&      41.68&      0.1623&    A&                2.42& 2.1&       \\
  sn2001V& Ia-norm&$0.93 \pm 0.05$&   28&     -15.50&      48.50&      0.2020&    A&                2.12& 4.4&       \\
\enddata
\tablecomments{
Overview of SN~Ia template spectra\citep{Matheson07}. Column (1) shows
the SN identifier. Column (2) shows the SN~Ia subtype, where
``Ia-91T'' indicate the overluminous, slow decliners, ``Ia-norm'' are
the ``normal'' SN~Ia, and ``Ia-91bg'' are the underluminous, fast
decliners. Column (3) shows the \dm15 of the SN~Ia. The number of
spectra $N$ are shown in Column (4), spanning a phase from $p_{min}$
to $p_{min}$ days (Column~(5)~and~(6)) with respect to the fitted
$V$-band peak.  Based on $w_{max}$ in Column (7) (The biggest weight
assigned to a the spectra for a given SNe), a grade $A$, $B$, or $C$
is assigned to each SNe (Column (8)), as described in
Section~\ref{sec:timeintspec}. The $\chi^2_f$ for the fit of the
time-integrated, dust-scattered, reddened, and flattened SN~Ia spectra
to the observed light echo spectrum is shown in Column (9). Column
(10) shows the $rlap$ value determined with \snid\ indicating the
correlation between the SN template and the observed light echo
spectrum.  An $rlap$ value bigger than 5 is considered a good
correlation.
}
\end{deluxetable}

\begin{deluxetable}{llrrrrcrrl}
\tabletypesize{\scriptsize} 
\tablecaption{
\label{tab:snoverviewIbc}}
\tablehead{
\colhead{SN} & \colhead{Subtype} & \colhead{N} & \colhead{$p_{min}$} & \colhead{$p_{max}$} & \colhead{$w_{max}$} & \colhead{Grade} & \colhead{$\chi^2_f*10^{-4}$} & \colhead{rlap} & \colhead{Ref.}  \\
\colhead{(1)} & \colhead{(2)} & \colhead{(3)} & \colhead{(4)} & \colhead{(5)} & \colhead{(6)} & \colhead{(7)} & \colhead{(8)} & \colhead{(9)} & \colhead{(10)} 
}
\startdata
  sn1994I& Ic-norm&   18&      -6.50&      63.50&      0.2703&    B&                2.90& 3.8&  \cite{Filippenko95}  \\
 sn1997ef&Ic-broad&   25&     -12.50&      82.50&      0.1485&    A&                4.08& 1.1&   \cite{Iwamoto00}    \\
 sn2004aw& Ic-norm&   25&      -7.50&      46.50&      0.1610&    A&                2.17& 3.7&   \cite{Taubenberger06}\\
 sn2005bf&  Ib-pec&   22&     -28.50&      33.50&      0.1474&    A&                6.68& 1.0&   \cite{Modjaz07}    \\
 sn2005hg& Ib-norm&   16&     -13.50&      26.50&      0.3285&    B&                7.45& 1.0&   \cite{Modjaz07}    \\
 sn2005mf& Ic-norm&    4&       3.50&      13.50&      0.4233&    C&                2.02& 4.2&   \cite{Modjaz07}    \\
\enddata
\tablecomments{
Same as Table~\ref{tab:snoverviewIa}, but with no \dm15 column,
Column~(2) shows the subtypes of SN~Ib/c, and Column~(10) shows the reference of the SN data.
}
\end{deluxetable}

\end{document}